\documentclass[11pt,letterpaper]{article}

\usepackage{amsmath,amsfonts,amssymb,eucal,graphicx,graphics,bm}
\usepackage[sort&compress,numbers]{natbib}
\usepackage{epsfig}
\usepackage{wrapfig}
\usepackage{subfigure}
\usepackage{amssymb}
\usepackage{epsfig}
\usepackage{subfigure}
\usepackage{graphicx}
\usepackage{color}
\usepackage[dvipsnames]{xcolor}
\usepackage{textcomp}
\usepackage{url}
\usepackage{float}
\usepackage{bm}
\usepackage{authblk}

\author[1]{Christopher P. Kempes}
\author[2]{Michael J. Follows}
\author[3]{Hillary Smith}
\author[4]{Heather Graham}
\author[3]{Christopher H. House}
\author[5]{Simon A. Levin}

\affil[1]{The Santa Fe Institute, Santa Fe, NM}
\affil[2]{Department of Earth, Atmospheric, and Planetary Sciences, Massachusetts Institute of Technology, Cambridge, MA}
\affil[3]{Department of Geosciences, Pennsylvania State University, University Park, PA}
\affil[4]{NASA Goddard Spaceflight Center, Greenbelt, MD}
\affil[5]{Department of Ecology and Evolutionary Biology, Princeton University, Princeton, NJ}

\title{Generalized Stoichiometry and Biogeochemistry for Astrobiological Applications}
\date{}

\begin{document}

\maketitle

\section*{Introduction}

A central need in the field of astrobiology is generalized perspectives on life that make it possible to differentiate abiotic and biotic chemical systems \cite{mckay2008approach}. A key component of many past and future astrobiological measurements is the elemental ratio of various samples. Classic work on Earth's oceans has shown that life displays a striking regularity in the ratio of elements as originally characterized by Redfield \cite{redfield1958biological,geider2002redfield,geoscience2014eighty}. The body of work since the original observations has connected this ratio with basic ecological dynamics and cell physiology, while also documenting the range of elemental ratios found in a variety of environments. Several key questions remain in considering how to best apply this knowledge to astrobiological contexts: How can the observed variation of the elemental ratios be more formally systematized using basic biological physiology and ecological or environmental dynamics? How can these elemental ratios be generalized beyond the life that we have observed on our own planet? Here we expand recently developed generalized physiological models \cite{kempes2012growth,kempes2016evolutionary,kempes2017drivers,kempes2019scales} to create a simple framework for predicting the variation of elemental ratios found in various environments. We then discuss further generalizing the physiology for astrobiological applications. Much of our theoretical treatment is designed for {\it in situ} measurements applicable to future planetary missions. We imagine scenarios where three measurements can be made --- particle/cell sizes, particle/cell stoichiometry, and fluid or environmental stoichiometry ---  and develop our theory in connection with these often deployed measurements. 

 Our general approach here is to focus on the macromolecules and physiology shared by all of life on Earth. For the macromolecules we are interested in components like proteins, nucleic acids, and cell membranes. For the shared physiology we consider processes such as growth rates, nutrient uptake, and nutrient storage, some of which are derivable from the macromolecular composition of cells. In thinking about the applicability of these two perspectives to life anywhere in the universe it is important to note that the specific set of macromolecules might vary significantly while the general physiological processes might be more conserved. However, our treatment of the macromolecules is easily generalized if one makes two assumptions: 1) that life elsewhere shares a set of macromolecules, even if that set is very different from Terran life, and 2) that those macromolecules fall along systematic scaling relationships. Throughout this paper we operate within these two assumptions and first address the observation and implications of (2), before moving on to a general treatment of physiological scaling which abstracts the underlying details of (1). Throughout we go back and forth between the patterns observed across single organisms of different size and the aggregate results for entire ecosystems composed of diverse organisms, which we characterize by a distribution of cell sizes. We discuss the general signatures of life that exist at both the cell and ecosystem level.

\section*{Deriving Elemental Ratios Across Cell Size}

Recently a variety of biological regularities have been discovered for life on Earth that show that organism physiology can be characterized by systematic trends across diverse organisms. These trends are often power-law relationships between organism size and a variety of physiological and metabolic features, and are derivable from a small set of physical and biological constraints \cite{kempes2019scales}. Intuitively, these relationships can be viewed as the  optimization of physiological function under fixed constraints through evolutionary processes \cite{kempes2019scales}. As such, in many contexts these scaling relationships may represent universal relationships connected to fundamental physical laws such as diffusive constraints. However, in many cases the cross-species scaling may reflect emergent and interconnected constraints of the physiology itself or of evolutionary history and contingency, in which case we might expect these scaling relationships to vary across life on diverse worlds.  For example, changes in the network architecture of the metabolism with size could be governed by the likelihood of cross-reactivity between molecules, which could depend on what types of molecules are being employed. In general, the possibility of contingent and emergent constraints is an important consideration for astrobiology.

Our interest here is in generalizing organism physiology and connecting it to stoichiometric ratio measurements that could be performed as part of astrobiological explorations of other planets using current or near-future instrumentation. Stoichiometry could be used as a relatively simple biosignature and, when considered within the context of the stoichiometry of the environment surrounding the particle/cell, could serve as a universal or ``agnostic'' biosignature. Agnostic biosignatures aim to identify patterns of living systems that may not necessarily share the same biochemical machinery as life on Earth. The need for reliable agnostic biosignatures increases as we examine planets deeper in the Solar System where common heritage with life on Earth is less likely. 

For the purposes of understanding elemental abundances, the cross-species trends in macromolecular components across the range of bacteria have been uncovered in recent work \cite{kempes2012growth,kempes2016evolutionary,kempes2017drivers}. For example, bacteria follow a systematic set of scaling relationships where protein concentrations are decreasing with increasing cell size and RNA components are increasing in concentration \cite{kempes2016evolutionary}. From the broad set of these relationships it is possible to derive the elemental ratio of a cell of a given size simply by considering the abundance and elemental composition of each component. The elemental ratio of the entire ecosystem is then found by considering the size distribution of organisms.

We calculate the total elemental abundances for a cell by knowing the elemental composition of a component, $c_{i}$ (e.g. N/protein), and the total quantity of that component, $n_{i}\left(V_{c}\right)$, in a cell of a given size $V_{c}$. The total abundance of one element, $E$ ($mol/cell$), is equal to the sum across all cellular components
\begin{equation}
E\left(V_{c}\right)=\sum_{i} c_{i}n_{i}\left(V_{c}\right),
\end{equation}
where the components are major categories of macromolecules such as proteins, ribosomes, and mRNA. Each of these components has a known scaling with cell size given in \ref{box-1}. As an example, the total nitrogen content in bacteria is given by 
\begin{equation}
E\left(V_{c}\right)=c_{N,p}n_{protein}+c_{N,DNA}n_{DNA}+c_{N,mRNA}n_{mRNA}+c_{N,tRNA}n_{tRNA}+c_{N,ribo}n_{ribosomes}+c_{N,l}n_{l}+c_{N,e}n_{e}
\end{equation}
where $c_{N,p}$ is the average number of N in protein, $c_{N,DNA}$, $c_{N,mRNA}$, $c_{N,tRNA}$, and $c_{N,ribo}$ are the average N in various types of DNA and RNA, $c_{N,l}$ is the N in lipids, and $c_{N,e}$ is the N in energy storage molecules such as ATP and carbohydrates. The counts of the macromolecules are given by $n_{protein}$, $n_{ribosomes}$, $n_{DNA}$,$n_{tRNA}$, $n_{mRNA}$, $n_{l}$, and $n_{e}$ which represent the numbers of proteins, ribosomes, DNA, tRNA, mRNA, lipids, and energy storage molecules in the cell, all of which depend on cell size (\ref{box-1}). For our analysis here we focus on N:P as an illustrative case, and thus typically ignore carbohydrates and lipids, which are minor cellular sources of these elements.

\begin{figure}
\centering
\includegraphics[width=.5\linewidth]{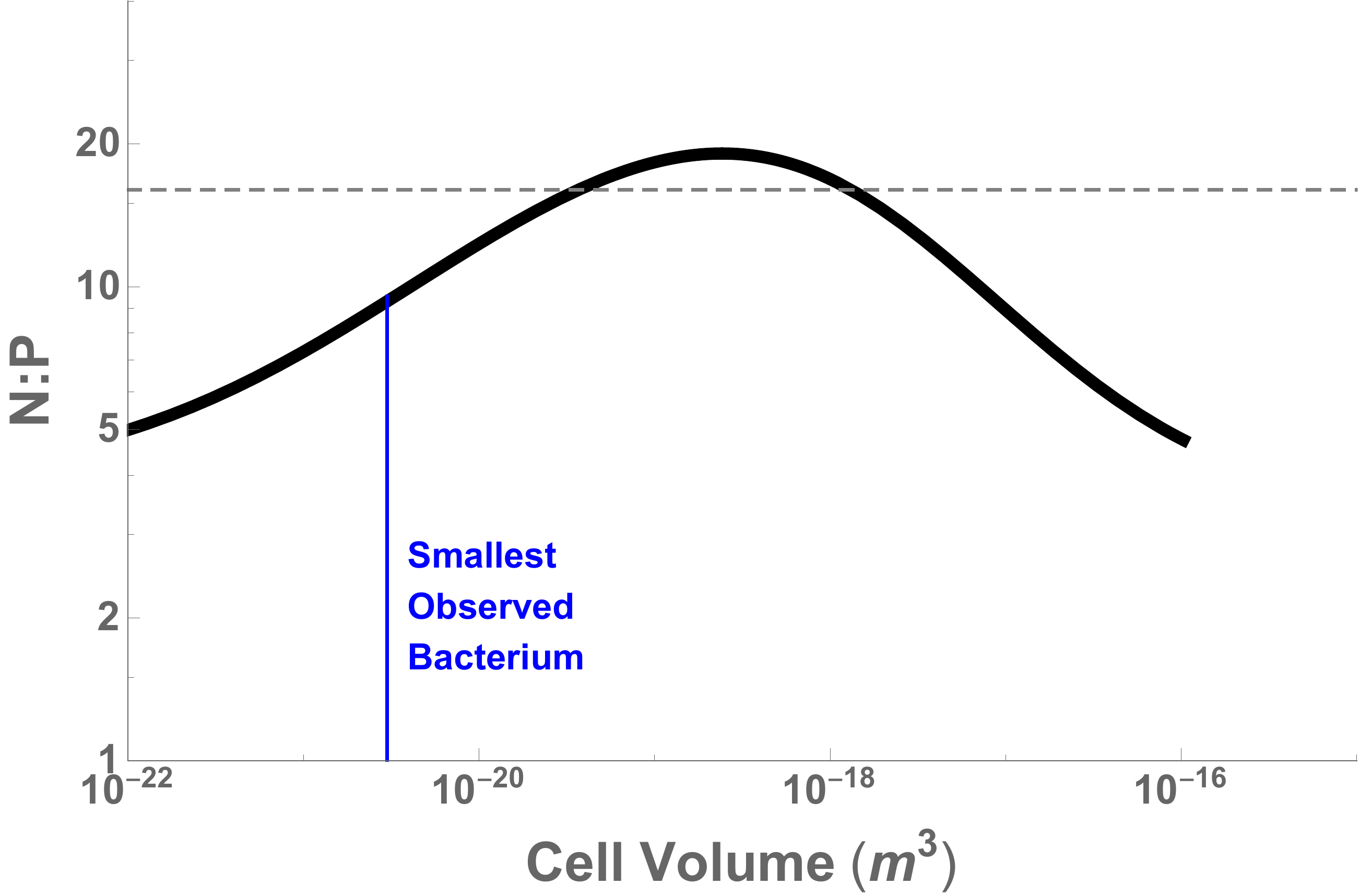}
\caption{Elemental ratios as a function of bacterial cell size showing a non-constant stoichiometry that often differs from the Redfield ratio (e.g. N:P of 16:1 indicated by the dashed line) for many cell sizes.}
\label{single-cell-redfield}
\end{figure}

Using typical values for the elemental composition of each component \cite{geider2002redfield}, Figure \ref{single-cell-redfield} gives the ratio of elements with overall cell size. This result shows that the elemental ratios agree with Redfield for some cell sizes but deviate significantly for most bacterial cell sizes. Both small and large bacterial cells have a decreased ratio of N to P compared with the Redfield ratio. It should be noted that the Redfield ratio is known to vary widely, and do so in ways that are ecologically meaningful from a resource competition perspective (e.g. \cite{geider2002redfield,klausmeier2004optimal,klausmeier2004phytoplankton,klausmeier2008phytoplankton}). We discuss these points in greater detail below.

These observations also show that one possible agnostic biosignature is non-constant elemental ratios across particle sizes. This is the result of evolution optimizing organism physiology at different scales \cite{kempes2019scales} which will lead to different ratios of macromolecules and thus different elemental ratios at each particle size. This is true even when the set of macromolecules is largely conserved across many species but the relative ratio of these macromolecules changes due to scaling laws with cell size, as is the case with ribosomes, DNA, and proteins. The strong and consistent trend of elemental ratios with cell size should be distinctly different from the patterns of abiotic particles. 

It should be noted that sampling issues may still exist. For example, it can be hard to separate biotic from abiotic particles in the Earth's oceans using known devices \cite{andersson1993proportion}. However, the stoichiometry of these particles once sorted are expected to radically differ, which should be systematically verified. Addressing these issues is an important topic of future work. In addition, it is important to note that these results are based on the macromolecular abundances of cells growing at maximum rate under optimal nutrient conditions, and cells are known to respond to environmental conditions by shifting macromolecular ratios and elemental abundances \cite{elrifi1985steady,healey1985interacting,rhee1978effects}. We address these processes of acclimation in our coupled biogeochemical model.

\begin{figure}[htb]
\includegraphics[width=.5\linewidth]{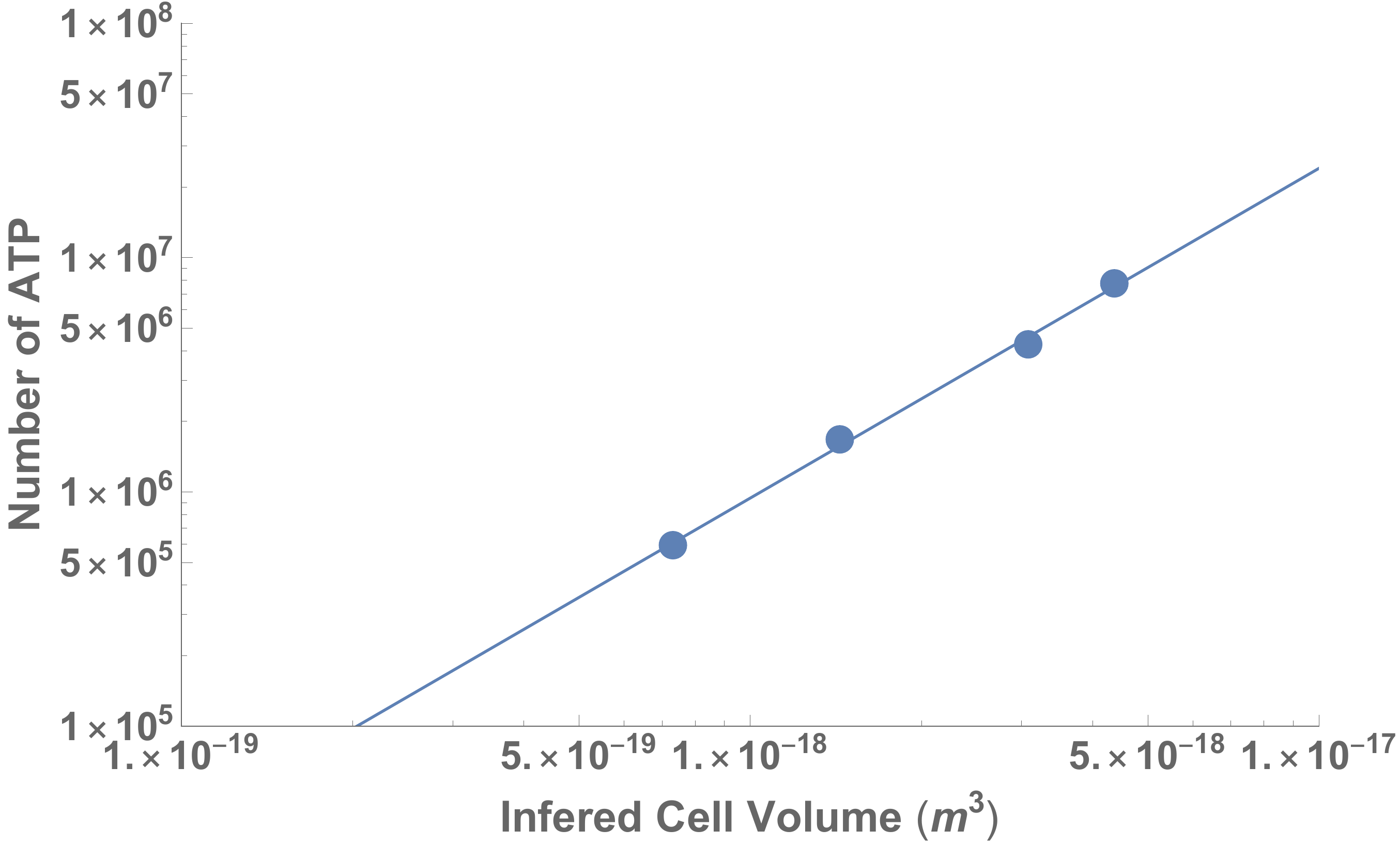}  
\caption{The number of ATP molecules as a function of cell volume in bacteria. The original data are from \cite{hamilton1967adenosine} where the original measurement of carbon content of a cell has been converted to cell volume using the relationship in \cite{lovdal}, and ATP mass per cell has been converted to counts per cell. The data follow $n_{a}=a_{0}V_{c}^{\beta_{a}}$ with $\beta_{a}=1.41\pm0.22$. }
\label{ATP-scaling}
\end{figure}

\newpage
\renewcommand{\thesection}{Box \arabic{section}}
\fcolorbox{black}{Apricot}{%
  \parbox{0.9\textwidth}{
  \footnotesize{
 \section{\small Equations governing macromolecular content in cells}
\label{box-1}
 
Many features of the cell have been previously shown to scale with overall cell size \cite{kempes2016evolutionary}. The scaling relationships for counts of the main macromolecular components follow
\begin{eqnarray}
n_{protein}&=&p_{0} V_{c}^{\beta_{p}} \\
\label{riboscaling}
n_{DNA}&=& d_{0}V_{c}^{\beta_{d}} \\
n_{ribo}&=& \frac{\bar{l}_{p} n_{protein} \left(\frac{\phi }{\mu} +1\right)}{\frac{\bar{r}_{r}}{\mu}-\bar{l}_{r} \left(\frac{\eta}{\mu}+1\right)} \\
n_{tRNA}&=& t_{0}n_{ribo}^{\beta_{t}} \\
n_{mRNA}&=& m_{0}n_{ribo}^{\beta_{m}} 
\end{eqnarray}
where $l_{r}$ is the average length of a ribosome in base pairs, $r_{r}$ (bp s$^{-1}$) is the maximum base pair processing rate of the ribosome which is assumed to be constant across both taxa and cell size, $\eta$ (s$^{-1}$) and $\phi$ (s$^{-1}$) are specific degradation rates for ribosomes and proteins respectively, and the $\mu$ is the growth rate of the cell. Some of these relationships are phenomenological, such as the scaling of protein content, while others can be derived from simple models. For example, the number of ribosomes is found using the 
coupled dynamics of protein and ribosome replication: 
\begin{eqnarray}
\frac{dn_{ribo}}{dt}&=&\gamma \frac{r_{r}}{l_{r}}n_{ribo}-\eta n_{ribo} \\
\frac{dn_{protein}}{dt}&=&\left(1-\gamma\right) \frac{r_{r}}{l_{p}}n_{ribo}-\phi n_{protein} 
\end{eqnarray}
where $\gamma$ is the fraction of ribosomes making ribosomal proteins. These equations can be solved analytically, where $\gamma$ can be found by enforcing that both the ribosomal and protein pools double at the same time, and Equation \ref{riboscaling} is given by the lifetime average of this solution. In addition to the average protein content of the cell, the ribosomal model also requires that we know the growth rate of an organism, which has also been shown to change with cell size \cite{kempes2012growth} based on the following simple model of energetic partitioning of total metabolism of a given cell size, $B_{0}V_{c}^{\beta_{B}}$, into growth and repair:
\begin{equation}
B_{0}V_{c}^{\beta_{B}}=E_{V}\frac{dV_{c}}{dt}+B_{V}V_{c}
\end{equation}
where $B_{V}$ (W m$^{-3}$) is unit maintenance metabolism, $E_{m}$ (J m$^{-3}$) is the unit cost of biosynthesis, $\beta_{B}\approx1.7$ is the scaling exponent of metabolic rate for bacteria, $B_{0}$ is a metabolic normalization constant with units W $(\text{m}^{3})^{-\beta_{B}}$. This equation can be solved for $V_{c}\left(t\right)$, the temporal growth trajectory of a cell, from which a time to reproduce can be found, which in turn gives the population growth rate as
\begin{equation}
\mu=\frac{\left(B_{V}/E_{V}\right)\left(1-\beta_{B}\right)\ln\left[\epsilon\right]}{\ln\left[\frac{1-\left(B_{V}/B_{0}\right)V_{c}^{1-\beta_{B}}}{1-\epsilon^{1-\beta_{B}}\left(B_{V}/B_{0}\right)V_{c}^{1-\beta_{B}}}\right]}.
\label{growth-pred}
\end{equation} 
This growth rate is found solving for the time to divide, $t_{d}$ in the equation $V_{c}\left(t_{d}\right)\equiv \epsilon V_{0}$, where $\epsilon\approx2$ is the ratio of the cell size at division compared to its initial size, $V_{0}$, and where $\mu=\ln\left(\mu\right)/t_{d}$. Finally, for the energy storage component of the macromolecular pool we should focus on ATP and ignore carbohydrates since we are concerned primarily with N:P ratios in this paper. The previous work cited above does provide scaling relationships or models for ATP, but Figure \ref{ATP-scaling} gives the dependence of total ATP on cell volume from data for marine bacteria \cite{hamilton1967adenosine} which is fit well by 
\begin{equation}
n_{a}=a_{0}V_{c}^{\beta_{a}}.
\end{equation}
It should be noted that some cells are known to have inorganic stores of phosphate and nitrate \cite{rhee1973continuous,galbraith2015simple}, and our treatment here does not account for such storage which is not characterized systematically for diverse bacteria.

}
  }%
}

\section*{Deriving Elemental Ratios in Environments From Size Distributions}

In many contexts it may be hard to precisely measure the changes in elemental ratios across a range of cell sizes and we may need to rely on bulk assessments of the entire environment, such as calculations for the total biomass as Redfield originally measured. Thus, it is useful to translate the above cell-level N:P ratios to whole-environment values. Here we will consider the value found from aggregating all particulate matter, later we will address both the aggregate particulate and surrounding fluid.

Given the strong connection between cell size and elemental ratios we can determine the aggregate elemental ratio within a microbial ecosystem by simply knowing the cell-size distribution. The total concentration of one element in an environment is given by 
\begin{equation}
E_{tot}=\int_{V_{min}}^{V_{max}}E\left(V_{c}\right)\mathcal{N}\left(V_{c}\right) dV_{c}
\label{total-element}
\end{equation}
where $\mathcal{N}\left(V_{c}\right)$ ($cells/m^{3}$) is the concentration of individuals of size $V_{c}$ in the environment (note that this equation holds for concentrations or frequencies of individuals), and $V_{min}$ and $V_{max}$ give the smallest and largest sizes respectively. 

\begin{figure}[h!]
\centering
\includegraphics[width=.45\linewidth]{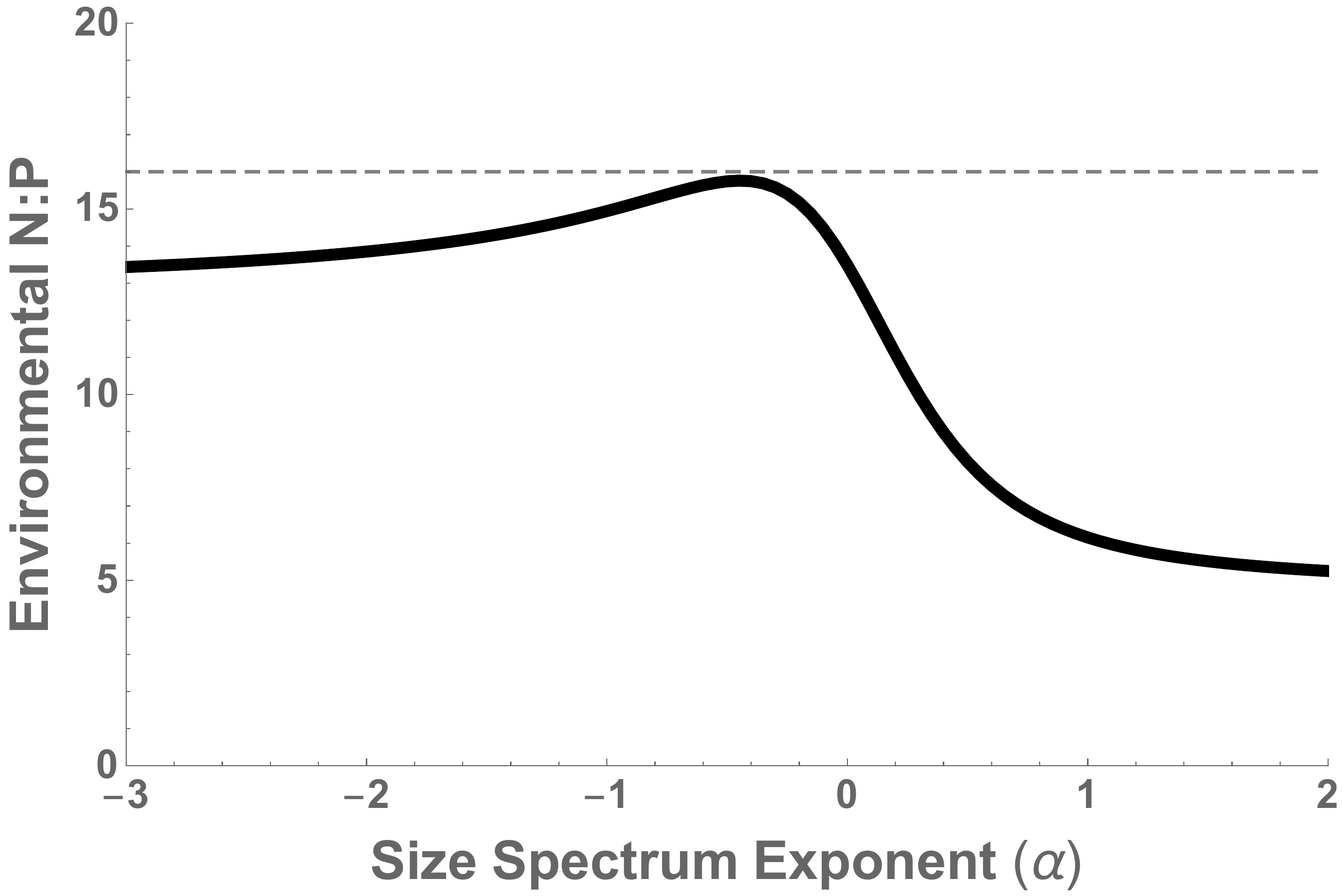}
\includegraphics[width=.45\linewidth]{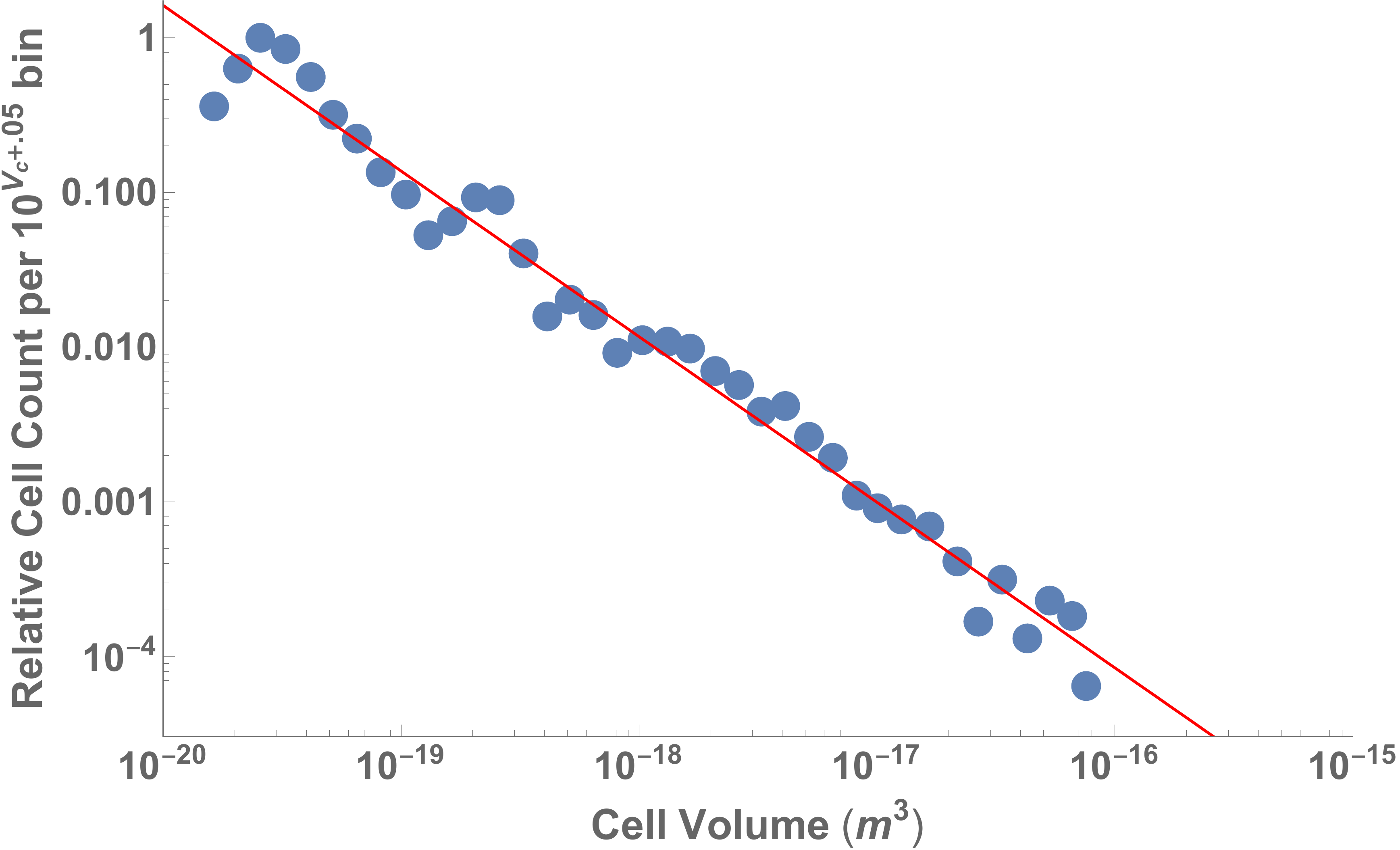}  
\caption{(a) Elemental ratios for an entire environment given a cell size distribution characterized by $f \propto V_{c}^{-\alpha}$ where $f$ is frequency and $V_{c}$ is cell size. The dashed lined is the standard Redfield N:P ratio of 16:1. For reference, (b) gives a measured size spectrum from \cite{cavender2001microbial} with a fitted exponent of $\alpha=-1.07\pm 0.05$. The comprehensive data from \cite{cavender2001microbial} show exponents that vary between $\alpha=-0.95$ and $\alpha=-1.35$.}
\label{environmental-redfield}
\end{figure}

To compare the elemental ratios we must first specify the frequency of individuals of different size. The distribution of individual sizes, often referred to as the size-spectrum, has been previously investigated in detail (e.g. \cite{sheldon1967continuous,cavender2001microbial,cuesta2018sheldon,ward2012size,taniguchi2014planktonic,irwin2006scaling}), and is observed to follow a variety of functional forms. One commonly observed relationship is a negative power law between cell size and abundance in an environment of the form $\mathcal{N}\left(V_{c}\right) = C V_{c}^{-\alpha}$, where \cite{cavender2001microbial} showed that exponents typically vary between $\alpha=-0.95$ and $\alpha=-1.35$ (see Figure \ref{environmental-redfield}b for an example abundance relationship).

Using Equation \ref{total-element} with the elemental relationships $E\left(V_{c}\right)$ from the previous section, and taking $\mathcal{N}\left(V_{c}\right) = C V_{c}^{-\alpha}$ we can explore the range of elemental ratios as a function of $\alpha$, where the value of $\alpha$ adjusts which cell sizes are being more heavily weighted in the integral. More specifically, $\alpha=0$ weights all cell sizes equally, more negative exponents increasingly weight smaller cells, and more positive exponents increasingly weight larger cells. In Figure \ref{environmental-redfield}a we have plotted the range of elemental ratios as a function of $\alpha$, where we find that only certain size distributions would produce values close to the typical Redfield ratio at the scale of an entire environment. Specifically, for $\alpha<0$, we find values that vary between 13:1 and 16:1 in the N:P ratio. The values most closely match the Redfield ratio of 16:1 for $\alpha=-1/2$, which differs slightly from the best fit exponent of $\alpha=-1.07\pm 0.05$ \cite{cavender2001microbial} (Figure \ref{environmental-redfield}b). However, it should be noted that characterizing the distributions of cell sizes as a power law is a simplification of more complicated distributions which often have a maximum abundance at an intermediate size \cite{sheldon1967continuous,cavender2001microbial}. The maximum abundance can be seen at the far left of Figure \ref{environmental-redfield}b where the peaked function is well approximated by a piecewise power law with a positive exponent on the left and negative exponent on the right. If we use the exact empirical function for $\mathcal{N}\left(V_{c}\right)$ over the range of bacterial sizes we calculate an N:P of $15.15$ which closely matches the Redfield ratio. 

These results show that our procedure generates {\it a priori} expectations of whole-environment stoichiometries from particle-size distributions and known organism physiology, and could be generalized to any distribution of cell sizes and any systematic physiology (e.g. presently unknown living systems that use a set of P and N-containing biomolecules different than Earth's proteins and nucleic acids). Even without generalizing physiology the variation in size distribution leads to a variety of total biomass N:P ratios. No single ratio can be relied on as a distinct biosignature. 

\section*{Generalized Physiological and Ecological Models of Biogeochemistry}

\begin{figure}[htb]
\includegraphics[width=.5\linewidth]{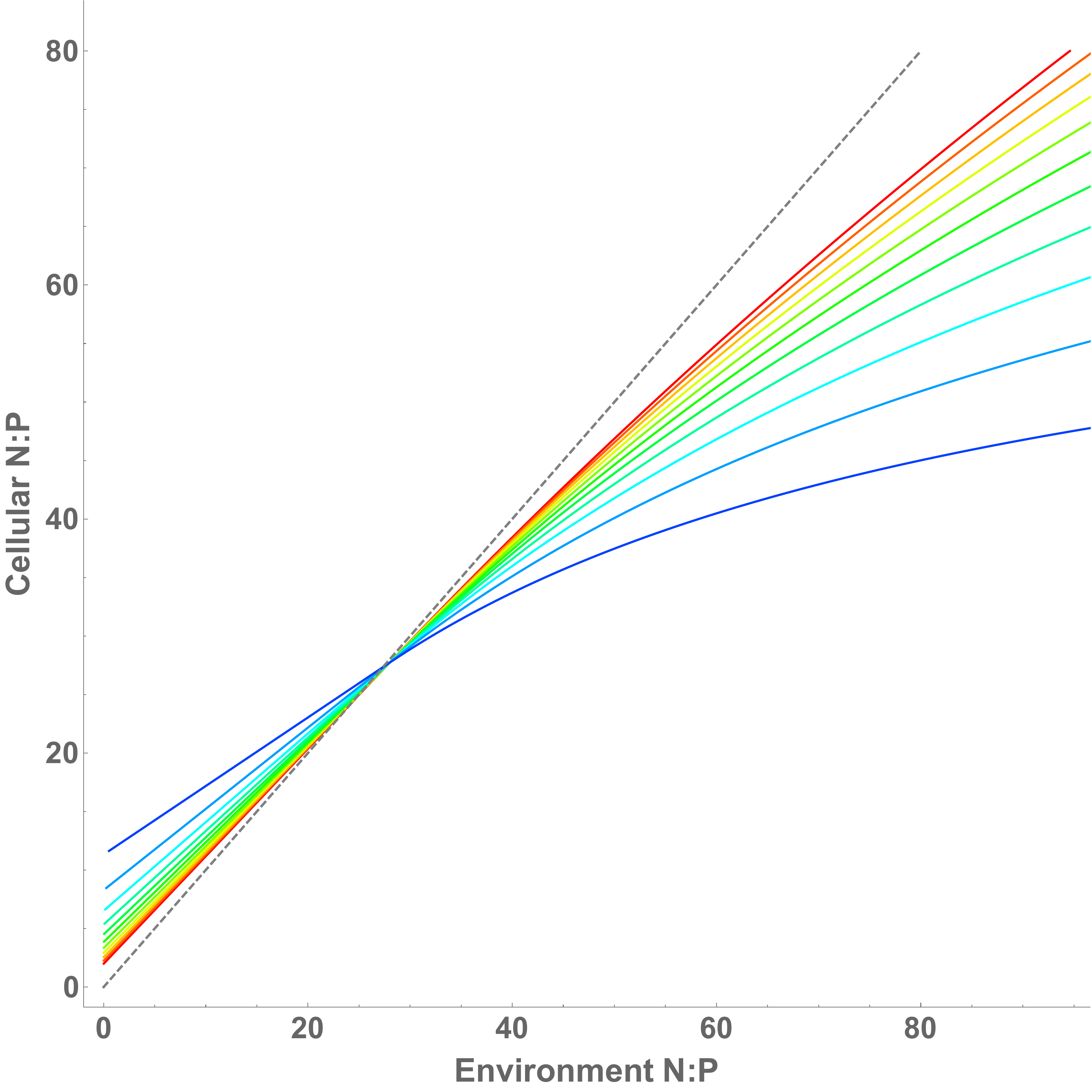}  
\includegraphics[width=.5\linewidth]{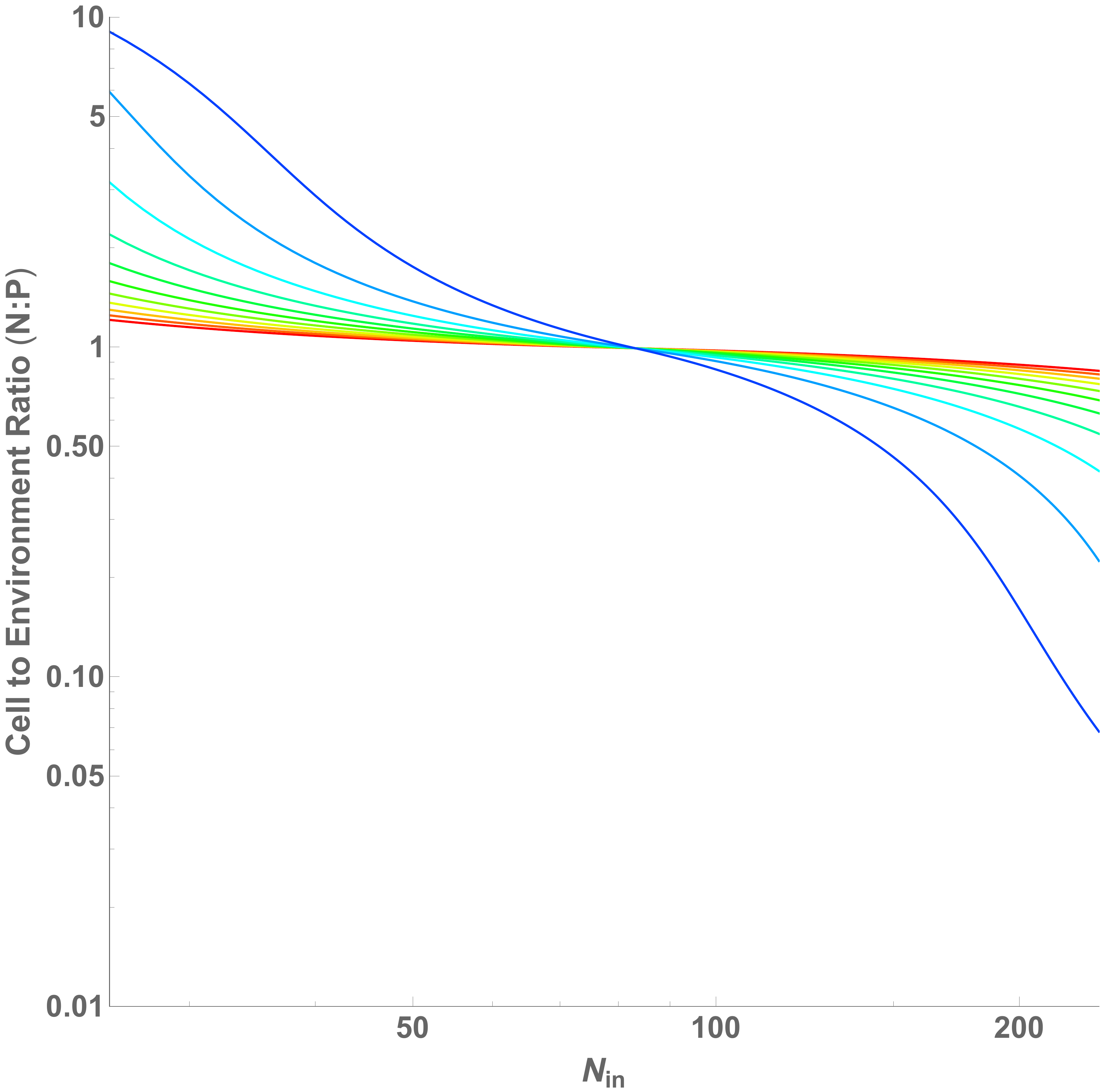}  
\caption{ (a) Elemental ratios within cells as a function of the environmental ratio and cell size (red is the smallest and blue is the largest cells), where the ecosystem is composed of only a single cell size. The dashed line is the one-to-one line. (b) The differences between cell stoichiometric ratios and the environment as a function of the nitrogen inflow, $N_{in}$, which is also the parameter being varied in (a).}
\label{cell-env-ratio}
\end{figure}

Thus far we have seen that strong trends in N:P with particle size could be an indicator of life, but that the total stoichiometric ratio of all biomass (filtered particles) does not have a single reliable value as a biosignature. We need more measurements to assess the ``livingness'' of a particular sample. One possibility is to simultaneously measure the particulate and environmental (fluid) stoichoiometries. It is also important to consider that macromolecular and elemental abundances in cells changes as cells acclimate to environmental constraints, where there are known physiological optima based on environmental conditions \cite{burmaster1979continuous,legovic1997model,klausmeier2004phytoplankton,klausmeier2004optimal,klausmeier2007model,klausmeier2008phytoplankton} and which is the topic of the following models.

A variety of efforts have shown how steady-state elemental ratios can be derived from physiological models coupled to flow rates in an environment \cite{legovic1997model,klausmeier2004phytoplankton,klausmeier2004optimal,klausmeier2007model,klausmeier2008phytoplankton}. These models are typically written as
\begin{eqnarray}
\frac{dR_{i}}{dt}&=&a\left(R_{i,0}-R_{i}\right)-f_{i}\left(R_{i}\right)\mathcal{N} \\
\frac{dQ_{i}}{dt}&=& f_{i}\left(R_{i}\right)-\mu(\vec{Q})Q_{i} \\
\frac{d\mathcal{N}}{dt}&=&\mu(\vec{Q})\mathcal{N}-m\mathcal{N}
\end{eqnarray}
where $\mu(\vec{Q})$ is the growth rate as a function of all of the existing elemental quotas (cellular quantities), and is typically given by 
\begin{equation}
\label{quotavec}
\mu(\vec{Q})=\mu_{\infty}\min \left(1-\frac{Q_{1,min}}{Q_{1}},1-\frac{Q_{2,min}}{Q_{2}},...,1-\frac{Q_{q,min}}{Q_{q}}\right)
\end{equation}
where $q$ is the total number of limiting elements \cite{legovic1997model,klausmeier2004phytoplankton,klausmeier2004optimal,klausmeier2007model,klausmeier2008phytoplankton}. The function $f_{i}\left(R_{i}\right)$ is the uptake rate for a given nutrient. The terms $Q_{i}$, $\mu_{\infty}$, and $f_{i}\left(R_{i}\right)$ are all known to systematically change with cell size (see \ref{box-2}), where commonly the uptake function is given by
 \begin{equation}
 f_{i}=U_{max} \frac{R_{i}}{K_{i}+r_{i}}
 \end{equation}
 given the half-saturation constant $K_{i}$ and the maximum uptake rate $U_{max}$ \cite{burmaster1979continuous}. In this model $a$ is the flow rate of the system, which affects both the inflow of nutrients from outside the system where $R_{i,0}$ is the concentration outside the system, and the loss of the nutrients from the system. Similarly, $m$ is the mortality rate of the cells and is often taken to be equal to the flow rate $a$ \cite{klausmeier2004phytoplankton,klausmeier2004optimal,klausmeier2007model,klausmeier2008phytoplankton}. In this system one nutrient is typically limiting because of the minimum taken in Equation \ref{quotavec}, and thus the equilibria of the system are typically dictated by the exhaustion and limitation of one nutrient. Previous work has shown that growth can be maximized in this framework by considering the allocation of resources to different cellular machinery, and that this leads to two optimum physiologies, one where maximum growth rate is optimized, and another where all of the resource equilibrium values are simultaneously minimized leading to resource colimitation and neutral competitiveness with all other species \cite{klausmeier2004phytoplankton,klausmeier2004optimal}.

In this model model the steady-state biomass, $\mathcal{N}^{*}$, limiting resource, $R^{*}$, and quota of the limiting resource, $Q^{*}$,  are given by 
\begin{eqnarray}
 \mathcal{N}^{*}&=& \frac{a\left(R_{in}-R^{*}\right)\left(\mu_{\infty}-m\right)}{Q_{min}\mu_{\infty}m}\\
R^{*}&=&\frac{Q_{min}m\mu_{\infty}K}{U_{max}\left(\mu_{\infty}-m\right)-Q_{min}\mu_{\infty}m} \\
Q^{*}&=& Q_{min}\frac{\mu_{\infty}}{\mu_{\infty}-m},
\end{eqnarray}
\cite{legovic1997model,klausmeier2004phytoplankton,klausmeier2004optimal,klausmeier2007model,klausmeier2008phytoplankton} where, for extant life, the physiological features are known to depend on size according to 
\begin{eqnarray}
U_{max}&=&U_{0}V_{c}^{\zeta} \\
\label{gen-phys-first}
K&=& K_{0}V_{c}^{\beta} \\
Q_{min}&=&Q_{0}V_{c}^{\gamma} \\
\mu_{\infty}&=&\mu_{0}V_{c}^{\eta}.
\label{gen-phys-last}
\end{eqnarray}
where the empirical values for the exponents and normalization constants are provided in \ref{box-2}. Given these general physiological scaling relationships the steady states are
\begin{eqnarray}
 \mathcal{N}^{*}\left(V_{c}\right)&=& \frac{a\left(R_{in}-R^{*}\right)\left(\mu_{0}V^{\eta}-m\right)}{m Q_{0}\mu_{0}V^{\gamma+\eta}}\\
R^{*}\left(V_{c}\right)&=&\frac{m Q_{0}\mu_{0}K_{0}V^{\gamma+\eta+\beta}}{U_{0}V^{\zeta}\left(\mu_{0}V^{\eta}-m\right)-mQ_{0}\mu_{0}V^{\gamma+\eta}} \\
Q^{*}\left(V_{c}\right)&=& \frac{Q_{0}\mu_{0}V^{\gamma+\eta}}{\mu_{0}V^{\eta}-m}
\end{eqnarray} 
where it is important to note that these equations provide results for a single cell size considered in isolation. Below we first consider how these functions change due to cell size using  known physiological scaling and then general exponents, and then we derive an ecosystem-level perspective from these results and discuss potential biosignatures under a range of exponent values.

\fcolorbox{black}{Apricot}{%
  \parbox{0.9\textwidth}{
  \small{
 \section{\normalsize Standard scaling relationships for physiological features}
\label{box-2}
A wide variety of organism features are known to depend on overall size for various taxa (e.g. \cite{andersen2016characteristic,brown2004tmt,west2005oas,savage2004pqp}), including the key features for biogeochemical considerations \cite{edwards2012allometric,litchman2007role,verdy2009optimal}. The physiological features of the coupled model are given by
\begin{center}
\begin{tabular}{ c | c | c | c}
& & Nitrogen & Phosphorous \\
\hline
 $U_{max}=U_{t}V_{c}^{\zeta_{t}}$ & $\zeta_{t}=0.67$ & $U_{t}=1.04\times10^{4}$ & $U_{t}=3.77\times10^{2}$ \\ 
 $K=K_{t}V_{c}^{\beta_{t}}$ & $\beta_{t}=0.27$ & $K_{t}=1.23\times10^{4}$ & $K_{t}=4.40\times10^{2}$ \\  
 $Q_{min}=Q_{t}V_{c}^{\gamma_{t}}$ & $\gamma_{t}=0.77$ & $Q_{t}=9.85\times10^{4}$ & $Q_{t}=3.56\times10^{3}$ \\
 $\mu_{\infty}=\mu_{t}V_{c}^{\eta_{t}}$ & $\eta_{t}=0.65$ & $\mu_{t}=4.02\times10^{12}$ & \\
\end{tabular}
\\
$\;$
\\
where the $t$ subscript indicates that these are the known values for extant Terran life.
\end{center}

}
  }%
}

\subsection*{Cell-level biogeochemistry for extant life}

First we consider the case where an environment is dominated by a single species, which would correspond to the measurement of a consistent particle size. Taking the known physiological scaling relationships for extant life (\ref{box-2}) we find that the size of the organism has a strong effect on the stoichiometric ratios of both the particles and fluid. Figure \ref{cell-env-ratio} gives the steady state N:P of cells as a function of steady state environmental N:P and cell size. We find that the smallest cells will show the greatest deviation from the environmental concentration for most environmental ratios. Differences between the fluid and particle stoichiometry may define a biosignature, and these will be most noticeable for environments dominated by the smallest cells. It should be noted that these results depend on the specific scaling relationships of the physiological features given in \ref{box-2}, which could greatly vary for life beyond Earth and are even known to vary across taxa for extant Terran life \cite{delong,kempes2012growth}.

\subsection*{Generalized ecosystem-level biogeochemistry}

Classic resource competition theory in equilibrium (e.g. \cite{tilman1982resource,levin1970community,hutchinson1953concept,hutchinson1957,volterra1927variazioni,volterra1931leccons}) indicates that for multiple species, in our case multiple cell sizes, to coexist on a single limiting resource they must all share the same $R^{*}$ value. This is not naturally the case given the physiological scaling relationships outlined above, or the unlikelihood that many species will have identical physiological parameter values. In general, at most $x$ number of species can coexist in equilibrium if there are $x$ independent limiting factors \cite{levin1970community}, and in our framework we can adjust the mortality rate, $m$, to abstractly represent the combination of many factors and to obtain coexistence. This adjustment could be the consequence of a variety of other factors such as variable predation, sinking rates, phage susceptibility, or intrinsic death. For our purposes this approach allows us to obtain a spectrum of cell sizes in connection with our earlier focus. 

To enforce coexistance we take $R^{*}\left(V_{c}\right)=R_{c}$, where $R_{c}$ is a constant, in which case the required mortality rate is given by 
\begin{eqnarray}
m&=&\frac{R_{c}U_{max}\mu_{\infty}}{R_{c}U_{max}+Q_{min}\mu_{\infty}\left(K+R_{c}\right)} \\
&=& \frac{R_{c}U_{0}\mu_{0}v^{\zeta+\eta}}{R_{c}U_{0}v^{\zeta}+Q_{0}\mu_{0}v^{\gamma+\eta}\left(R_{c}+K_{0}v^{\beta}\right)}.
\end{eqnarray}
This function for $m$ should be seen as the consequence of the complicated evolutionary dynamics of many species living in a coupled ecosystem where prey and predator traits have evolved over time and new effective niches have emerged. It should also be noted $m$ is now size dependent compared with being set to constant value which was the case for the earlier results.

Our mortality relationship can be incorporated into $\mathcal{N}^{*}$ to give the scaling of biomass concentration for each cell size:
\begin{equation}
\mathcal{N}^{*} \left(V_{c}\right)=\frac{a\left(R_{in}-R_{c}\right)V_{c}^{-\zeta}\left(R_{c}+K_{0}v^{\beta}\right)}{R_{c}U_{0}}.
\label{n-scaling}
\end{equation}
This result has two important limits, where either the half-saturation constant is much smaller than the equilibrium value of nutrient in the environment, $K_{0}v^{\beta}\ll R_{c}$, or is much bigger than this environmental concentration, which leads to
\begin{equation}
\mathcal{N}^{*}\left(v\right)=
\begin{cases} 
     \propto V_{c}^{-\zeta} & K_{0}V_{c}^{\beta}\ll R_{c} \\
      
     \propto V_{c}^{\beta-\zeta} & K_{0}V_{c}^{\beta}\gg R_{c} 
   \end{cases}
   \label{n-cases}
   \end{equation}
These two relationships provide nice bounds on the scaling of $\mathcal{N}$ given the underlying physiological dependencies. 

Similarly, the quota is given by 
\begin{equation}
Q^{*}=Q_{0}V_{c}^{\gamma}+\frac{R_{c}v^{\zeta-\eta}U_{0}}{\mu_{0}\left(R_{c}+K_{0}v^{\beta}\right)}.
\end{equation}
which implies that the ratio of particle to fluid elemental abundance for the limiting nutrient is the following function of cell size
\begin{equation}
\frac{\mathcal{N}^{*}Q^{*}}{R^{*}}=
\begin{cases} 
     \frac{a (R_{in}-R_{c}) \left(Q_{0} V_{c}^{\gamma -\zeta}+\frac{U_{0}
   V_{c}^{-\eta }}{\mu_{0}}\right)}{R_{c} U_{0}} & K_{0}V_{c}^{\beta}\ll R_{c} \\
      
    \frac{a (R_{in}-R_{c}) \left(K_{0} Q_{0} \mu_{0}
   V_{c}^{\beta +\gamma -\zeta }+R_{c} U_{0} V_{c}^{-\eta }\right)}{
  R_{c}^2 U_{0}\mu_{0}} & K_{0}V_{c}^{\beta}\gg R_{c}
   \end{cases}
   \label{n-cases}
   \end{equation}
This relationship is similar to the types of results shown in Figure \ref{cell-env-ratio}, but gives the ratio between cell and environment concentrations for a single element of interest (rather than as comparisons of ratios of elements), and importantly, does so under the constraints of coexistence. This result leads to particular biosignature possibilities when measuring only a single element, and does so for the more realistic ecosystem conditions of coexistence. If we measure the particle size distribution in an environment, then this is enough to specify the value of $\alpha=-\zeta$ or $\alpha=\beta-\zeta$ from Equation \ref{n-cases}, leaving us with $\gamma$ and $\eta$ to determine the element ratio scaling between cells and the environment as a function of particle size. 

From a biosignatures perspective, the most ambiguous measurement would be particles that perfectly mirror the environmental stoichiometry where $N^{*}Q^{*}/R^{*}$ equals a constant for all particle sizes. In the first limit, $K_{0}V_{c}^{\beta}\ll R_{c}$, this would require $\zeta=\gamma=-\alpha$ and $\eta=0$. This result would imply that the quota and uptake rates would need to scale with the same exponent and as the negative value of the size exponent, both of which are consistent with the observations of \ref{box-2} and Figure \ref{environmental-redfield}b for extant life. However, this result also requires that there would be no change in growth rate with cell size, which is very unlikely from a variety of biophysical arguments.

In the second limit, $K_{0}V_{c}^{\beta}\gg R_{c}$, a constant value of $N^{*}Q^{*}/R^{*}$ requires that $\zeta-\beta=\gamma=-\alpha$ and $\eta=0$. Again the absence of changes in growth rate connected with $\eta=0$ is unlikely. In addition, under this scenario the difference in the uptake and half-saturation scaling, represented by $\zeta-\beta$, must equal the scaling of the quota and take the opposite value as the size-spectrum scaling, which is a combination that is not consistent with extant life and is a very special case in general. Thus, under both limits $N^{*}Q^{*}/R^{*}$ is unlikely to have a constant value as a function of cell size, and an observed scaling in this ratio forms a likely biosignature. 

\begin{figure}[h!]
\includegraphics[width=.9\linewidth]{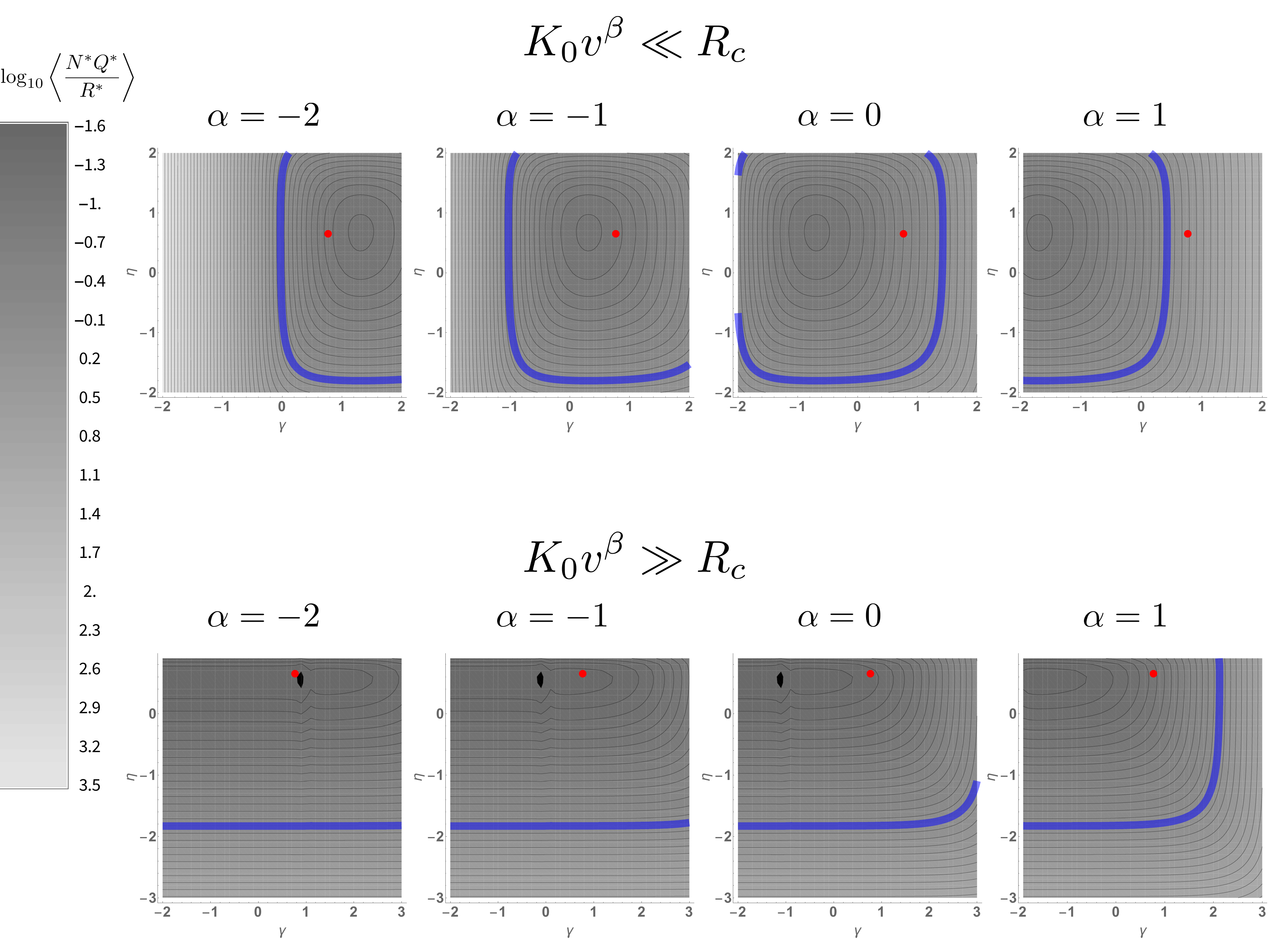}  
\caption{The ratio of the cellular to environmental nitrogen, $\left\langle \mathcal{N}^{*} Q^{*}/R^{*}\right\rangle$, as function of the size spectrum exponent $\alpha$, the minimum quota (cellular requirement) scaling exponent, $\gamma$, and the growth rate scaling exponent, $\eta$. We have shown the results for the two approximations $K_{0}v^{\beta}\ll R_{c}$ and $K_{0}v^{\beta}\gg R_{c}$. In each plot the blue line represents $\left\langle \mathcal{N}^{*} Q^{*}/R^{*}\right\rangle=1$, or $\log_{10}\left\langle \mathcal{N}^{*} Q^{*}/R^{*}\right\rangle=0$. The red point represents the known $\gamma$ and $\eta$ exponent values for extant life from \ref{box-2}.}
\label{exponent-ratios}
\end{figure}

This potential biosignature still requires one to measure the cell-size spectrum in detail, which may be challenging in certain settings or with certain devices. However, these relationships can be easily translated into an aggregate ecosystem-level measurement by averaging over all coexisting cells, where the average is given by 
\begin{eqnarray}
\left\langle \frac{\mathcal{N}^{*} Q^{*}}{R^{*}}\right\rangle&=& \frac{1}{V_{max}-V_{min}}\int_{V_{min}}^{V_{max}}\frac{\mathcal{N}^{*}\left(V\right)Q^{*}\left(V\right)}{R^{*}}dV \\
= && \left.\frac{a\left(R_{in}-R_{c}\right)V_{c}\left(\frac{R_{c}U_{0}V_{c}^{-\eta}}{1-\eta}+\frac{R_{c}Q_{0}\mu_{0}V_{c}^{\gamma-\zeta}}{1+\gamma-\zeta}+\frac{Q_{0}K_{0}\mu_{0}V_{c}^{\beta+\gamma-\zeta}}{1+\beta+\gamma-\zeta}\right)}{\left(V_{max}-V_{min}\right)R_{c}^{2}U_{0}\mu_{0}}\right\rvert_{V_{c}=V_{min}}^{V_{c}=V_{max}}
\end{eqnarray}
which, considering the two approximations for $\mathcal{N}$, becomes 
\begin{equation}
\left\langle \frac{\mathcal{N}^{*} Q^{*}}{R^{*}}\right\rangle=
\begin{cases} 
      \left.\frac{a\left(R_{in}-R_{c}\right)V_{c}\left(\frac{U_{0}V^{-\eta}}{\mu_{0}\left(1-\eta\right)}+\frac{Q_{0}V_{c}^{\gamma-\zeta}}{1+\gamma-\zeta}\right)}{\left(V_{max}-V_{min}\right)R_{c}U_{0}}\right\rvert_{V_{c}=V_{min}}^{V_{c}=V_{max}} & K_{0}V_{c}^{\beta}\ll R_{c} \\
      
      \left.\frac{a\left(R_{in}-R_{c}\right)V_{c}\left(\frac{R_{c}U_{0}V_{c}^{-\eta}}{1-\eta}+\frac{Q_{0}K_{0}\mu_{0}V_{c}^{\beta+\gamma-\zeta}}{1+\beta+\gamma-\zeta}\right)}{\left(V_{max}-V_{min}\right)R_{c}^{2}U_{0}\mu_{0}}\right\rvert_{V_{c}=V_{min}}^{V_{c}=V_{max}} & K_{0}V_{c}^{\beta}\gg R_{c}  
   \end{cases}
   \end{equation}
To fully specify this community level ratio for generalized life we need to constrain the normalizations constants, $Q_{0}$, $k_{0}$, $U_{0}$, and $\mu_{0}$ given any choice of the exponents. A reasonable way to determine the values of these constants is to match the generalized rates to the observed Terran rates from \ref{box-2} at a particular reference size, $V_{r}$, which leads to
\begin{eqnarray}
U_{0} &=& U_{t}V_{r}^{\zeta_{t}-\zeta} \\
K_{0} &=& K_{t}V_{r}^{\beta_{t}-\beta} \\
Q_{0} &=& Q_{t}V_{r}^{\gamma_{t}-\gamma}\\
\mu_{0} &=& \mu_{t}V_{r}^{\eta_{t}-\eta}.
\end{eqnarray} 
After calibrating the constants to an intermediate cell size of $V_{r}=10^{-18}$ (m$^{3}$), Figure \ref{exponent-ratios} gives the community level $\left\langle \mathcal{N}^{*} Q^{*}/R^{*}\right\rangle$ as a function of the scaling exponents. When $K_{0}v^{\beta}\ll R_{c}$ the size exponent $\alpha$ specifies $-\zeta$, and when $K_{0}v^{\beta}\gg R_{c}$ then $\alpha$ specifies $\beta-\zeta$. In both approximations we plot $\left\langle \mathcal{N}^{*} Q^{*}/R^{*}\right\rangle$ as a function of $\eta$ and $\gamma$ for a range of $\alpha$ values.

We find that typically the $\left\langle \mathcal{N}^{*} Q^{*}/R^{*}\right\rangle$ differs from $1$ for a wide range of $\alpha$, $\eta$ and $\gamma$ values. This is true under both limits. For fixed values of $\alpha$ the $\left\langle \mathcal{N}^{*} Q^{*}/R^{*}\right\rangle=1$ line is a closed curve as a function of $\eta$ and $\gamma$ (Figure \ref{exponent-ratios}). This curve defines the regime within which it is possible to find $\left\langle \mathcal{N}^{*} Q^{*}/R^{*}\right\rangle=1$ for any value of either $\eta$ and $\gamma$, and this region covers a wide range of exponent values. However, the known values of $\eta$ and $\gamma$ for extant life occur fairly far from this curve and would show elemental concentrations that are distinguishable from the environment. It is likely that the full range of $\alpha$, $\eta$ and $\gamma$ combinations explored here are precluded for biophysical reasons, but this requires more detailed work in the future. Finally, it is important to note that most $\alpha$, $\eta$ and $\gamma$ combinations would yield cell to environment ratios that significantly differ from $1$, and that the gradients are very steep around the $\left\langle \mathcal{N}^{*} Q^{*}/R^{*}\right\rangle$=1 line. Thus, it is a fairly safe assumption that the elemental abundances of cells should differ from the environment as this would be the expectation for physiological scaling chosen at random. 

\fcolorbox{black}{Apricot}{%
  \parbox{1.0\textwidth}{
  \small{
 \section{\normalsize Summary of potential biosignatures}
\label{box-3}
 \begin{enumerate}
\item Systematic shifts in stochiometry with particle size
\item Particle sizes that follow a power law distribution for abundance
\item Systematic shifts in the ratio of particle to fluid elemental abundance as a function of particle size
\end{enumerate}
}
  }%
}

\section*{Discussion}
The general framework provided here should make it possible to assess biosignatures for a wide diversity of potential life (see Box 3 for a summary). We focused on bacterial life as an example of what we would expect in ecosystems dominated by the simplest life. However, all of our results could be tuned to other classes of organisms with the appropriate changes in scaling relationships for macromolecular content and abundance distributions. Our results do this generally for any life which is governed by a set of physiological scaling relationships, where, for example, the nutrient quotas are abstracting the underlying changes in macromolecules and could represent a diverse set of alternate physiologies and sets of macromolecules for alternate evolutionary histories or origins of life. The main assumption in this generalized physiological model is that life will fall along allometric scaling relationships, which has good justification from various arguments connected with universal physical constraints. In addition, it should be noted that many of the physiological scaling relationships have strong physical principles motivating the exponents and the wide variation taken in the generalized equations may not be realizable by life anywhere in the universe. Thus observed biosignatures may be much more similar to our analyses in Figures \ref{single-cell-redfield}, \ref{environmental-redfield}, and \ref{cell-env-ratio} than the possibilities encapsulated in our generalized physiological model as capture in Figure \ref{exponent-ratios}. 

In addition, our efforts here have often focused on the assumption of one limiting nutrient. However, this scenario of a single resource typically does not lead to coexistence (e.g. \cite{tilman1982resource,levin1970community,hutchinson1953concept,hutchinson1957,volterra1927variazioni,volterra1931leccons}). The problem of coexistence can be solved by many additional considerations such as environmental stochasticity, the addition of spatial dynamics, or species adaptation (e.g. \cite{hutchinson1953concept,levins1971regional,klausmeier2002spatial,kremer2013coexistence}), all of which could be important for future modeling efforts or for measurements of the spatial variation in stoichometry. However, our solution for mortality allows for coexistence in a single environment and our model is compatible with measurements made at a single or coarse-grained location which may be typical of many future astrobiological measurements. Our general physiological perspective should be combined with more advanced biogeochemical models that consider many nutrients, including trace elements, and more complex ecological and evolutionary dynamics --- many of which can be connected systematically with size \cite{andersen2016characteristic,kempes2019scales} --- to fully explore the range of particle size distributions, and the particle to fluid stoichiometric differences that can be reasonably expected to represent biosignatures.

\section*{Acknowledgements} 
The authors thank Natalie Grefenstette for useful discussions and comments. This work was supported by a grant from the Simons Foundation (\#395890, Simon Levin), and grants from the National Aeronautics Space Administration (80NSSC18K1140, Christopher Kempes, Hillary Smith, Heather Graham, and Christopher House), and the National Science Foundation (OCE-1848576, Simon Levin).

\bibliography{scaling-refs}

\end{document}